\documentclass[12pt]{article}
\usepackage{graphics}
\usepackage{amssymb}
\usepackage{psfrag}

\newcommand{\rf}[1]{(\ref{#1})}
\newcommand{\beq}{\begin{equation}}
\newcommand{\eeq}{\end{equation}}
\newcommand{\bea}{\begin{eqnarray}}
\newcommand{\eea}{\end{eqnarray}}

\newcommand{\sg}{\sigma}

\newcommand{\mi}{\!-\!}
\newcommand{\equ}{\!=\!}

\begin{document}

{\normalsize \hfill SPIN-05/28}\\
\vspace{-1.5cm}
{\normalsize \hfill ITP-UU-05/34}\\
${}$\\

\begin{center}
\vspace{48pt}
{ \Large \bf The Universe from Scratch}

\vspace{30pt}

{\sl J. Ambj\o rn}$\,^{a,c}$,
{\sl J. Jurkiewicz}$\,^{b}$
and {\sl R. Loll}$\,^{c}$

\vspace{24pt}
{\footnotesize

$^a$~The Niels Bohr Institute, Copenhagen University\\
Blegdamsvej 17, DK-2100 Copenhagen \O , Denmark.\\
{ email: ambjorn@nbi.dk}\\

\vspace{10pt}

$^b$~Mark Kac Complex Systems Research Centre,\\
Marian Smoluchowski Institute of Physics, Jagellonian University,\\
Reymonta 4, PL 30-059 Krakow, Poland.\\
{email: jurkiewicz@th.if.uj.edu.pl}\\

\vspace{10pt}

$^c$~Institute for Theoretical Physics, Utrecht University, \\
Leuvenlaan 4, NL-3584 CE Utrecht, The Netherlands.\\
{email:  j.ambjorn@phys.uu.nl, r.loll@phys.uu.nl}\\

\vspace{10pt}

}
\vspace{28pt}

\end{center}

\begin{center}
{\bf Abstract}
\end{center}
\noindent
A fascinating and deep question about nature is what one would see if one 
could probe space and time at smaller and smaller distances. Already the 
19th-century founders of modern geometry contemplated the possibility that a 
piece of empty space that looks completely smooth and structureless to the 
naked eye might have an intricate microstructure at a much smaller scale. 
Our vastly increased understanding of the physical world acquired during the 
20th century has made this a certainty. The laws of quantum theory tell us that
looking at spacetime at ever smaller scales requires ever larger energies, and, 
according to Einstein's theory of general relativity, this will alter spacetime itself: 
it will acquire structure in the form of ÒcurvatureÓ. What we still lack is a definitive
{\it theory of quantum gravity} to give us a detailed 
and quantitative description of the highly curved and quantum-fluctuating 
geometry of spacetime at this so-called Planck scale. --
This article outlines a particular approach to constructing such a theory, that
of {\it Causal Dynamical Triangulations}, and its achievements so far in
deriving from first principles why spacetime is what it is, from the tiniest realms
of the quantum to the large-scale structure of the universe.  
\vspace{12pt}
\noindent


\newpage

\subsection*{Searching for the quanta of spacetime}

Armed with last century's insights into the nature of both quantum theory and general relativity, 
physicists believe that probing the structure of space and time at distances far below those 
currently accessible by our most powerful accelerators would reveal a rich geometric fabric, 
where spacetime itself never stands still but instead quantum-fluctuates wildly. One of the 
biggest challenges of theoretical physics today is to identify these fundamental ``atoms" or 
excitations of spacetime geometry and understand how their interaction gives rise to the 
macroscopic spacetime we see around us and which serves as a backdrop for all known 
physical phenomena. 

Two pillars of contemporary physics support the expectation that as we resolve the fabric of 
spacetime with an imaginary microscope at ever smaller scales, spacetime will turn from an 
immutable stage into the actor itself. First, due to Heisenberg's uncertainty relations, probing 
spacetime at very short distances is necessarily accompanied by large quantum fluctuations 
in energy and momentum - the shorter the distance, the larger the energy-momentum 
uncertainty. Second, according to Einstein's theory of general relativity, the presence of these 
energy fluctuations, like that of any form of energy, will deform the geometry of the spacetime 
in which it resides, imparting curvature which is detectable through the bending of light rays 
and particle trajectories. Taking these two things together leads to the prediction that the 
quantum structure of space and time at the so-called Planck scale must be highly curved and 
dynamical. 

A long held ambition of theoretical physicists is to find a consistent description of this dynamical 
microstructure within a {\it theory of quantum gravity}, which unifies quantum theory and general 
relativity, and to determine its ramifications for high-energy physics and cosmology. Given the 
extraordinary smallness of the Planck length, how can we achieve progress in describing a 
physical situation that cannot be directly probed by experiment in the foreseeable future? The 
way this is usually done is by first postulating additional dynamical principles or fundamental 
symmetries at small distances, which are not accessible to direct experimental verification, 
second, verifying that these do not conflict with standard quantum physics or general relativity 
as one goes to larger scales, and third, predicting new physical phenomena that can (at least 
in principle) be tested, or confirmed indirectly by astrophysical observations. Examples of 
fundamental building principles are that the universe is made up of tiny vibrating strings, or that 
spacetime at the Planck scale is not a continuum, but consists of tiny discrete grains. 

Research into quantum gravity falls broadly into two categories \cite{kiefer,roads}: 
nonperturbative approaches to quantum gravity, whose primary aim is to quantize the 
gravitational degrees of freedom per se, introducing little or no additional structure such 
as supersymmetry or extra dimensions, and 
string-theoretic approaches, where 
the quantization of gravity appears almost as a by-product of a unified higher-dimensional 
and supersymmetric ``theory of everything", whose fundamental objects are strings and 
(mem)branes \cite{witten,strings}.
 
The research program that will be described in this article deals with the investigation of causal 
nonperturbative quantum gravity and belongs in the first category. The approach takes its name 
from the main technical tool it employs to try and construct a theory of quantum gravity, namely, 
{\it Causal Dynamical Triangulations}, or CDT for short.\footnote{Previous reviews of
CDT, describing the general theory and covering earlier results in lower dimensions can
be found in \cite{previous}.} What makes this approach particularly
interesting is the fact that it has recently produced a number of tangible results which mark it as
a serious contender for the still elusive theory of 
quantum gravity \cite{ajl-prl,semi,spectral,universe}. Firstly, there is evidence that
the theory has a good classical limit. This means that it reproduces Einstein's classical theory
at sufficiently large scales: when one ``zooms out" the imaginary microscope from 
the scale at which the quantum fluctuations take place, one eventually rediscovers the smooth
four-dimensional spacetime of general relativity. Secondly, we have first indications of what
the quantum structure of spacetime may be at the Planck scale.

We will explain in the following why these results are indeed remarkable and how they were
obtained, in a manner hopefully accessible to those outside the field. The emphasis will be on 
describing the (very few) fundamental building principles that go into the construction of the
theory, on explaining the main results and their physical significance, and on giving an idea
of where we are headed. Before doing this, we will in the next section set out by sketching 
some of the problems facing research in quantum gravity, in order to provide the reader with
a better idea of how the results described later appear in a larger context.

\subsection*{Why quantum gravity is special}

Quantum gravity is quite unlike any other
fundamental quantum interaction in that it describes the dynamics of an entity
that in most physical situations is considered as fixed and given, namely, that 
of spacetime itself. Recall that the degrees of freedom of a spacetime in classical 
general relativity can be described by the spacetime metric $g_{\mu\nu}(x)$, 
a local field variable which determines
the values of distance and angle measurements in spacetime, or, equivalently,
how spacetime is bent and curved locally.\footnote{When using the term (quantum)
spacetime, we will in the following always mean the abstract spacetime
(a differential manifold in the classical case) {\it together} with its metric properties.}
What spacetime {\it is} classically is
determined by solving the Einstein equations for $g_{\mu\nu}(x)$, subject to
boundary conditions and a particular matter content of the universe or a piece
thereof. In the
same manner, in order to determine what spacetime {\it is} from a 
quantum-theoretical point of view, one would like to formulate a quantum
analogue of Einstein's equations, from which ``quantum spacetime" should 
then emerge as a {\it solution}. 

This should be contrasted with usual quantum field theory,
which describes the dynamics of elementary particles
and their interactions on a {\it fixed} spacetime background, usually 
that of the flat, four-dimensional Minkowski space of special relativity. 
Since at short distances\footnote{Short, but still far away from the Planck scale
defined below.} the gravitational forces are so much weaker
than the electromagnetic ones, say, it is usually an excellent approximation
to treat the gravitational degrees of freedom as ``frozen in" and 
non-dynamical. The trivial geometric structure of the
Minkowski metric forms merely part of the immutable background structure of
how quantum field theories are formulated. On the other hand, the physical
situations that quantum gravity aims to explain are not in general  
describable in terms of linear fluctuations of the metric field around 
Minkowski space or some other fixed background metric. These include
the quantitative description of ``empty" spacetime at very short distances
of the order of the Planck scale, $10^{-35}$ m, and of the extreme and 
ultradense state our universe presumably was in when it was very young.   
From a technical point of view this implies that in quantum gravity one 
has to modify standard quantization techniques which rely (sometimes
implicitly) on the presence of a fixed metric background structure. This is
often phrased by saying that gravity must ultimately be quantized in a way that
is both {\it background-independent} (i.e. does not distinguish any particular
background metric at the outset) and {\it nonperturbative} (i.e. does not
simply describe the dynamics of linear perturbations around some fixed 
background spacetime). 

Decades of quantum gravity research, including numerous trials and errors, 
have convinced many of the necessity of background-independence and
a nonperturbative quantization approach \cite{smolin}, and
the last twenty years or so have seen intense efforts to develop
alternative ways of quantizing applicable to the case of a dynamical 
spacetime geometry. Because of the complicated mathematical structures
involved, this turns out to be very difficult. Singularly unhelpful in this
endeavour has been the absence of experimental or observational data 
to guide the search for the correct theory of quantum gravity. This happens
because the extreme scales that are necessarily involved in physical
situations where quantum-gravitational effects are important are 
not accessible directly with current technology. A possible strategy to
address this state of affairs is to take a rather conservative approach to
theory-building, for example, to avoid going out on a limb by postulating 
the existence
of new physical quantities and symmetries for which there is as yet no
evidence. As we will see, the approach of CDT is fairly minimalist in
that it takes a set of well-known physical principles and tools
(quantum-mechanical superposition, causality, triangulation of geometry,
elements of the theory of critical phenomena) and merely adapts them to the situation of
a dynamical geometry.   

How far then have we got in our quest for a theory of quantum gravity?
Maybe the best answer to this question is that it is hard to tell, as long
as we do not know the final answer with some certainty. This 
assessment must in all honesty also
include the possibility that we are still very far from the correct theory. There has
been a lot of research since the late 1980s, both in nonperturbative 
quantum gravity as such and under the heading of string theory,
at the time when the incorporation of gravitation into our understanding of the
quantum world emerged as a final theoretical frontier of 
fundamental physics. Progress has undoubtedly been made,
at the very least in terms of developing technical tools to
describe quantum geometry nonperturbatively (see, for example,
\cite{thie,rovelli,4drev}).
However, in order to cut a long story short, there is still {\it not a single 
theory} of
quantum gravity that is both reasonably complete and internally
consistent mathematically. By reasonably complete we
mean that it should provide answers to some of quantum
gravity's central questions, for instance, ``Why is spacetime the way it is?",
``What are the fundamental excitations of quantum geometry?",
``What are the quantum properties of black holes?" etc.,
even if they are not immediately verifiable experimentally.
Our entire discussion therefore must be understood 
against a background where we do not have 
a plethora of ``possible" theories available
(and just look for clues for how to pick the right one), but where we
are still looking for the first instance of a quantum gravity theory
that is sufficiently complete to make at least some predictions
about the quantum behaviour of spacetime. 

\subsection*{The dynamical principle underlying CDT}

The most important theoretical tool of the CDT method to construct 
a quantum theory of gravity is Feynman's principle of superposing quantum
amplitudes \cite{path}, the famous {\it path integral}, applied to spacetime
geometries. Its basic idea, familiar from quantum mechanics, is to
obtain a solution to the quantum dynamics of a physical system
by taking a superposition of ``all possible" configurations of the
system, where each configuration contributes
a complex weight $\exp (iS)$ to the path integral, which depends on
the classical action $S=\int dt L(t)$ of the configuration, where $L$ denotes the
system's Lagrangian. For the
case of a nonrelativistic particle moving in a potential, the configurations
are literally paths in space, i.e. continuous trajectories ${\bf x}(\tau)$ describing
the particle's position as a function of time $\tau$, which runs through
an interval $\tau\in [0,t]$. Superposing (that is, adding or integrating up) the associated
quantum amplitudes $\exp iS^{\rm part}[{\bf x}(\tau)]$ as in
eq.\ \rf{prop} below, one obtains a solution to the
Schr\"odinger equation of the particle.
\begin{figure}[t]
\centerline{\scalebox{0.55}{\rotatebox{0}
{\includegraphics{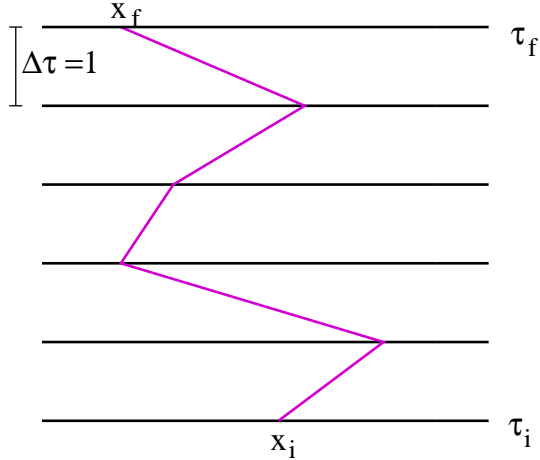}}}}
\caption[particle]{\small A typical, piecewise straight path 
${\bf x}(\tau)$ contributing to
the regularized Feynman path integral for a non-relativistic particle
moving between two points ${\bf x}_{\rm i}$ 
and ${\bf x}_{\rm f}$.
}
\label{particle}
\end{figure}
 It is important to realize that
the {\it individual} paths ${\bf x}(\tau)$ appearing in the path integral are {\it not}
themselves physical trajectories the particle could move on, and even
less solutions to the particle's classical equations of motion.
Instead, they are so-called ``virtual" paths, that is, a bunch of curves
one can draw between fixed initial and final points ${\bf x}_{\rm i}$ 
and ${\bf x}_{\rm f}$ (Fig.\ \ref{particle}). The magic of the path integral 
\begin{equation}
G({\bf x}_{\rm i},{\bf x}_{\rm f},t):=\;\;\;\sum\hspace{-1.2cm}
\int\limits_{ {\rm paths:}\, {\bf x}_{\rm i}\rightarrow
{\bf x}_{\rm f}} {\rm e}^{iS^{\rm part}[{\bf x}(\tau)]}
\label{prop}
\end{equation}
is that the true quantum physics of the particle is encoded precisely in the
{\it superposition} of all these virtual paths\footnote{In using the
notation $\Sigma\hspace{-.3cm}\int$ in \rf{prop}, we want to indicate that
the path integral may be a sum or a genuine integral (or possibly
a combination of both), depending on whether the configurations
contributing to it are labelled by discrete or continuous 
parameters. In CDT we will meet an example of the former.}. 
In order to extract these
physical properties, one has to evaluate suitable quantum
operators $\hat O$ on the ensemble of paths contributing to
\rf{prop}. For example, one may be interested in
computing expectation values for the position or the energy of the
particle, together with their quantum fluctuations. Of course, the path 
integral or ``propagator" \rf{prop} also allows us to retrieve the classical 
behaviour of the particle in a particular limit (in this case, when its mass
becomes big), but it contains more information, describing the full quantum 
dynamics of the system.

Analogously, a path integral for gravity is a superposition of all virtual ``paths" 
our universe (or a part thereof) can follow as time unfolds. These paths are simply
the different configurations for the metric field variables $g_{\mu\nu}(x)$
mentioned earlier.\footnote{We are considering here only the gravitational
degrees of freedom, that is, a path integral for ``pure gravity", and thus a
``theory for empty space". Matter fields can in principle be included without
problems.} 
 It is important to realize that a single path is now no
longer an assignment of just three numbers (the coordinates $x_i$ of the
particle) to every moment $\tau$ in time, but rather the assignment to
every $\tau$ of a whole array of numbers (the
space-space components ${\bf g}_{ij}(x)\equiv {\bf g}_{ij}({\bf x},\tau)$ of the metric
tensor $g_{\mu\nu}(x)$) {\it for each spatial point} ${\bf x}$.
This is simply a consequence of gravity being a field theory with
infinitely many degrees of freedom. The path integral for gravity can thus
be written as
\begin{equation}
G({\bf g}_{\rm i},{\bf g}_{\rm f},t):=\;\;\;\sum\hspace{-1.5cm}
\int\limits_{ {\rm spacetimes:}\, {\bf g}_{\rm i}\rightarrow
{\bf g}_{\rm f}} {\rm e}^{iS^{\rm grav}[g_{\mu\nu}({\bf x},\tau)]},
\label{propgrav}
\end{equation}
where $S^{\rm grav}$ now denotes the classical gravitational action 
associated with a spacetime metric $g_{\mu\nu}$ with initial and
final boundary condition ${\bf g}_{\rm i}$ and ${\bf g}_{\rm f}$, separated
by a time distance $t$. Like in
the particle case, the individual spacetime configurations interpolating
between the initial and final spatial geometries have nothing a priori to
do with classical spacetimes, and are much more general objects. 
Again, one would expect to be able to retrieve the full quantum dynamics of
spacetime from the path integral \rf{propgrav}, which is a superposition
of all possible ways in which an empty spacetime can be curved\footnote{These
are also sometimes called ``spacetime histories".}. In
other words, the collective behaviour of the virtual spacetimes contributing
to the gravitational propagator \rf{propgrav} should tell us what
quantum spacetime {\it is}. To extract this geometric information, we will
again have to evaluate suitable quantum operators $\hat O$ on the
ensemble of geometries contributing to \rf{propgrav}. Suffice it to say
that making the gravitational path integral well-defined and extracting the 
desired physical information is very much more difficult than in
the case of the quantum particle.

The way in which CDT proceeds is by giving a precise prescription of
how the path integral \rf{propgrav} should be computed, and in particular
how the class of virtual paths should be chosen. In addition, it provides a
set of technical tools to extract concrete physical information about
the quantum geometry thus created by the principle of quantum 
superposition.

There are a number of ways in which the path integral of CDT differs from
that of previous approaches. In the first instance, it is genuinely
nonperturbative, in that the contributing geometries can have very large
curvature fluctuations at very small scales and thus be arbitrarily far
away from any classical spacetime. Our summation is ``democratic" in
that no particular spacetime geometry is distinguished at the outset.
In fact, path integral histories which have any geometric resemblance to a 
classical spacetime are so rare that their contribution to the path integral
is effectively negligible.\footnote{This is completely analogous to the particle
case, where it can be shown rigorously that classical paths ``form a set of
measure zero" with respect to the Wiener measure of the 
path integral \cite{wiener}. Maybe surprisingly, the paths which contribute
non-trivially are nowhere differentiable, and thus ``consist only of
corners". One expects a similarly nonclassical behaviour for the
dominant configurations of the gravitational path integral.} 
Secondly, as we will see in the following, the
causal structure of the geometries contributing to the path sum plays
an important role in the method of causal dynamical triangulations,
and is a key new element in comparison with previous, so-called
Euclidean path-integral approaches to quantum gravity.

\subsection*{Representing spacetime geometry in CDT}

What we need to do next in order to make sense of the 
expression \rf{propgrav} for the nonperturbative quantum-gravitational 
propagator is to define the
precise class of spacetime geometries (labelled above by $g_{\mu\nu}$) 
over which the sum or integral is to be taken. As elsewhere in quantum
field theory, one is immediately confronted with the fact that unless one
chooses a careful regularization for the path integral, it will be wildly
divergent and simply not exist in any meaningful mathematical sense
(and thus be useless for extracting physical information).
``Regularizing" means making the path integral finite by introducing 
certain cutoff parameters for the contributing configurations, which
at a later stage will be removed in a controlled manner.

Before defining a suitable class of geometries in the next section, we will
first explain the nature of the regularized spacetimes used by CDT,
which are called
``piecewise flat geometries". Recall that the dynamical degrees of
freedom of a geometry are the ways in which it is locally curved. 
Piecewise flat geometries are simply spaces that are flat (the same as
straight or uncurved, that is, structureless from a geometric point of
view) everywhere apart from small subspaces where curvature is
said to be concentrated. This in a way discretizes curvature and
vastly reduces the different number of ways spacetime can be curved. The type
of geometry we will use is a triangulated space, also sometimes called a 
Regge geometry, after the physicist who first introduced it into
(classical) general relativity \cite{regge}. It can be thought of as a space glued
together from elementary building blocks which are (higher-dimensional
generalizations of) triangles, so-called ``simplices". The geometric
structure of each simplex is trivial, since it is by itself flat by definition 
and therefore carries no curvature. Local curvature only appears 
along lower-dimensional interfaces when 
one starts gluing the simplices together. 

\begin{figure}[t]
\psfrag{(a)}{{\bf{\LARGE (a)}}}
\psfrag{(b)}{{\bf{\LARGE (b)}}}
\centerline{\scalebox{0.5}{\rotatebox{0}{\includegraphics{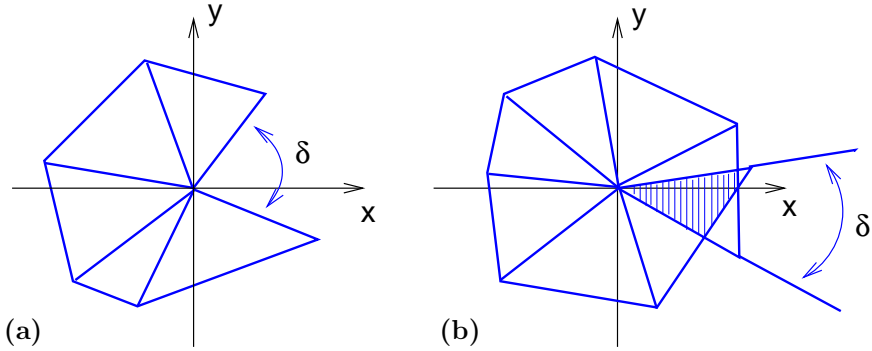}}}}
\caption[euangle]{\small Example of positive and negative deficit angles
$\delta$
located at a vertex of a two-dimensional triangulation. The triangulations
have been cut open in order to flatten them into the two-dimensional
Euclidean $x$-$y$-plane. To reconstruct the original geometry
around the central vertex (which has been placed at the plane's origin),
one has to identify the triangle edges as indicated by the arrows. In case
(a), the angles meeting at the vertex add up to less than 2$\pi$, resulting
in a positively curved space upon regluing. In case (b) the combined
angle exceeds 2$\pi$, corresponding to a negatively curved space.}
\label{euangle}
\end{figure}
This can be visualized most
easily in the two-dimensional case. Consider a set of identical
equilateral two-dimensional
triangles cut out from a piece of cardboard which is perfectly straight and
unbendable (and hence flat). To obtain a larger surface, start gluing these
triangles together by identifying their one-dimensional sides or edges pairwise.  
Points where several edges meet are also called vertices. One can obtain a
piece of flat space by arranging the triangles in a regular pattern so
that exactly six triangles and edges meet at each vertex. However,
there are many more ways to create {\it curved} spaces by the same
gluing procedure. Namely, whenever the number of triangles meeting
at a vertex is smaller or larger than six, this vertex will carry a positive
or negative curvature.\footnote{Equivalently, one speaks of the vertex
having a positive or negative deficit angle, simply because the sum of the
angles of the triangles contributing at the vertex is smaller or bigger
than $2\pi$, whereas in the flat case it is exactly $2\pi$, see Fig.\ \ref{euangle} for
illustration.}
By ``curvature" we mean the {\it intrinsic} curvature
of the two-dimensional surface, i.e. the curvature that can be detected
from within the surface -- for example, by studying the trajectories of
particles or light rays --, and is independent of any higher-dimensional
space in which it may be imbedded. This mirrors a property of the
physical theory of general relativity 
in four dimensions, which likewise depends only on the
intrinsic geometry of spacetime. The set-up in higher dimensions
is identical, with the two-dimensional triangles (or ``two-simplices") 
substituted by 
the corresponding flat higher-dimensional simplices (three-simplices
(or tetrahedra) in
dimension 3, four-simplices in dimension 4, etc.). 
Generally speaking, the fundamental building blocks in 
dimension $d$ are
glued together pairwise along their $(d\mi 1)$-dimensional faces, and
their intrinsic curvature is concentrated on the $(d\mi 2)$-dimensional
intersections of these faces. 

The so-called Regge calculus \cite{reggereview} 
was originally designed to approximate 
smooth classical
spacetimes, or, more precisely, solutions to the Einstein equations, 
by these piecewise flat, triangulated spaces. There are two reasons
for why this is a very economical way of describing a spacetime. Firstly,
only a finite amount of data is necessary to completely characterize
a finite piece of spacetime, namely, the geodesic invariant lengths 
of all the one-dimensional
edges of all the simplices involved, and the way in which
the $d$-dimensional simplices are glued together.\footnote{This can
be kept track of by attaching labels to all simplices and their faces, say,
and then pairing off the faces which are glued together.}
Secondly, because no coordinate system need ever be introduced
on the simplices, this formulation does not share the usual
coordinate redundance of Einstein gravity described in terms of the
field variables $g_{\mu\nu}(x)$. (The latter overcount the
degrees of freedom, because $g_{\mu\nu}$'s which are related by
coordinate transformations correspond to one and the same
physical geometry.)

The use of Regge geometries in the quantum theory is not new,
and CDT builds on previous attempts of both ``Quantum Regge
Calculus" \cite{hamber} and ``Dynamical Triangulations" \cite{edt} 
to define a theory
of quantum gravity from a nonperturbative Euclidean path 
integral\footnote{The essential difference between the two approaches
is that in Quantum Regge Calculus one fixes a triangulation or
``gluing", so that the path integral takes the form of a (multiple)
integral over the lengths of the edges of that triangulation, whereas
in Dynamical Triangulations one fixes {\it all} edge lengths to a common 
value
$a$, in which case the path integral takes the form of a discrete sum
over all inequivalent ways to glue the (then identical-looking)
simplicial building blocks
together.\label{diff}}. 
To avoid misunderstandings, it should be emphasized that the use
of triangulated spacetimes differs in classical and quantum applications.
The objective in the former is to approximate a single, smooth spacetime
(which may or may not be known exactly by some other method) 
as well as possible. This can be achieved by choosing a sequence of
triangulations, where in each step of the sequence 
the triangulation is chosen finer
than in the previous step
(i.e. the typical edge length is decreased) and therefore can
converge to the smooth manifold in a pointwise sense. 
In the two-dimensional example, it is quite clear that such an
approximation can be very good when the edge lengths become much smaller
than the scale at which the smooth spacetime is curved.   

By contrast, the objective of the quantum theory, and that of CDT in
particular, is to approximate
{\it the integral} \rf{propgrav} as well as possible, or, more precisely, to 
{\it define} it since there is currently no alternative, independent
way of doing the computation. This is a completely different task,
since the integral does not represent a single classical geometry,
but a quantum superposition, where each single 
contributing spacetime is a highly nonclassical object, as we
pointed out earlier. There is no accurate mathematical statement
to guide this construction, but one would expect that the
path integral should provide an ``ergodic sampling" of the space of
geometries. This may seem like a very vague characterization,
but one is in practice very much constrained by the requirement of
making the regularized path integral mathematically well-defined
and obtaining a sensible classical limit.     

The short-distance
cutoff $a$ is an important part of our regularization of the spacetime
geometries in the gravitational propagator. We will take the limit $a\rightarrow
0$ as part of the search for a so-called continuum limit of the
path integral over the regularized geometries. This has to be done
in order to obtain a final theory which does not depend on
many of the arbitrary details which have gone
into constructing the regularized model, which itself constitutes only
an intermediate step in the construction of the theory. 
Using a finite ``lattice spacing" $a$ and taking 
$a\rightarrow 0$ 
(while renormalizing the coupling constants of the theory as a
function of $a$) 
is a method
borrowed from the theory of critical phenomena and virtually
ensures that the end result does not depend on a variety of
regularization details. This latter property of ``universality" is
only a necessary condition 
and does by no means guarantee that this construction leads
to a viable theory of quantum gravity, as opposed to describing
the dynamics of certain one-dimensional polymers, say,
as we will explain further in the next section.\footnote{One 
may feel tempted to postulate that the cutoff
$a$ is a fundamental shortest length scale pertaining to the
existing physical world, and thus do away with any continuum limits.
Apart from having to justify such an ad-hoc assumption,
one would then have to face the fact that the quantum processes 
at this scale, which such a ``theory" may describe,
will by default depend on a large number of a priori arbitrary 
regularization parameters labelling all possible choices of 
fundamental building blocks and gluing rules, 
and thus run the danger of having no predictive value at all.}

\subsection*{The ensemble of virtual spacetime geometries in CDT}

Now that we have introduced the regularized triangulated geometries
the question still remains as to exactly what ensemble of such
objects should be included in the sum over geometries in \rf{propgrav}.
Here is where CDT differs in a crucial way from previous approaches,
and where the notion of ``causality" comes into play. We mentioned
above that precursors of CDT's nonperturbative path integral are
``Euclidean" in nature. What this means is that the integration is not
performed over so-called Lorentzian {\it spacetimes} (which have one time-
and three space-directions) but over Euclidean {\it spaces} (which
have four spatial directions, and thus no notion of time, light rays or
causality). Classically, Euclidean ``spacetimes" are bizarre and
unphysical entities, in which moving back and forth in time is just as
easy as moving back and forth in space. Their use in the 
(mainly perturbative) gravitational path integral 
was made popular in the late 1970s by the influential work of S. Hawking 
and collaborators on black holes and quantum cosmology in 
the context of Euclidean quantum gravity \cite{eqg}. The reason for using
them instead of Lorentzian spacetimes of the correct physical 
signature\footnote{The {\it signature} refers to the signs of the
four eigenvalues of the symmetric matrix $g_{\mu\nu}(x)$; it is
(+,+,+,+) in the Euclidean case and (-,+,+,+) in the Lorentzian case.}
is mainly technical: in the Euclidean case, the
weights $\exp (iS^{\rm grav})$ are no longer complex but real numbers,
which simplifies a discussion of the convergence properties of
the path integral, and also makes Monte-Carlo simulations
possible.\footnote{In order to simplify notation, we will always use
the notation $\exp (iS)$ to denote Boltzmann weights, with the implicit
understanding that $S$ is a real action when we talk about Lorentzian
signature (and the weight thus a complex phase factor), and a purely
imaginary one in a Euclidean context (and $\exp (iS)$ therefore a
{\it real} quantity).\label{boltz}} The potential catch is that in gravity, unlike
in other quantum field theories on a flat background, there is no
obvious relation between a nonperturbative path integral for
Lorentzian and one for Euclidean geometries. In fact, causal
dynamically triangulated gravity in dimensions two \cite{2dcdt}, 
three \cite{3dcdt} and four \cite{ajl1,ajl4d,ajl-prl,semi,spectral,universe,
blackhole} has for the first time provided
concrete evidence that the two path integrals are genuinely
inequivalent and possess completely different properties.
The remainder of this section explains in more detail how the
path integral for either Euclidean or Lorentzian gravity 
in terms of triangulated geometries is set up and
evaluated, and at the same time retraces some of the history
that led to the introduction of CDT.  

With the ingredients
that were defined in the previous section, it would seem straightforward
to write down a regularized version of the gravitational propagator
as
\begin{equation}
G^{\rm reg}({\bf T}_{\rm i},{\bf T}_{\rm f},t):=\;\;\;\sum_{ {\rm triangulations\, T:}
\, {\bf T}_{\rm i}\rightarrow
{\bf T}_{\rm f}} {\rm e}^{iS^{\rm reg}[T]},
\label{propreg}
\end{equation}  
where $T$ denotes a triangulated spacetime, glued from four-simplices, and
with two spatial triangulated boundary geometries ${\bf T}_{\rm i}$ and
${\bf T}_{\rm f}$ (glued from three-simpli\-ces), between which it interpolates.
The gravitational action for a piecewise flat spacetime $T$
schematically takes the form
\begin{equation}
S^{\rm reg}(T)= -1/G_N\,{\it Curvature}(T)+\lambda\,{\it Volume}(T),
\label{regact}
\end{equation}
and there is a definite prescription for how to compute the curvature 
and volume of a given triangulation $T$ in terms of the lengths of its
edges and its connectivity (that is, the way the four-simplices are glued
together).
The two coupling constants of the theory appearing in \rf{regact} are
Newton's constant $G_N$, governing the strength of gravitational 
interactions, and the cosmological constant $\lambda$, another constant
of nature, which may be responsible for the ``dark energy" 
pervading our universe \cite{sahni}.

As mentioned in footnote \ref{diff}, all simplices used in DT are 
equilateral\footnote{To be precise, this is true for the Euclidean version
of dynamical triangulations (DT); CDT operates with {\it two} different
edge lengths, one for all edges that have a spatial orientation, and
one for edges with a time-like orientation \cite{ajl4d}.}, and the path integral
assumes the form of a discrete sum over inequivalent ways in
which the simplicial building blocks can be glued together.
The only thing that remains then to be specified in \rf{propreg}
is whether any gluing of the building blocks is to be allowed, or
whether further restrictions need to be imposed. One condition turns
out to be essential for making the path integral construction 
well-defined. Call ${\cal N}(N_4)$ the number of distinct gluings
of $N_4$ four-simplices, for a particular set of gluing rules.
Clearly, this number will grow with $N_4$, but the important
question is whether it will grow exponentially as a function of
$N_4$ or faster, namely, ``super-exponentially", for example,
like $\exp (cN_4^\nu)$, with $c>0$ and 
$\nu >1$. In the latter case, and noting
that $N_4(T)$ is proportional to {\it Volume(T)}, there is no way
in which the exponential weights $\exp {iS^{\rm reg}[T]}$ 
could ever counterbalance the growth of the number of
contributing geometries (the growth of the so-called ``entropy" of the
system). The path integral would then be too divergent to
lead to a fundamental theory of gravity. 

These considerations preclude the inclusion in the path
integral \rf{propreg} of a so-called ``sum over topologies".\footnote{The
{\it topology} of a spacetime describes the way in which it hangs together.
For example, the topology of a two-dimensional compact and closed
surface is completely characterized by the number of its ``holes" or ``handles".
It could have the form of a surface of a ball (no holes), of a surface of
a torus or bicycle inner tube (one hole), of a surface of a double-torus
or two-hole doughnut
(two holes), and so on. In four dimensions, the labelling of different topologies
is much more involved.} Therefore, the topology of the spacetimes
contributing to the nonperturbative path integral has to be fixed. It is
typically chosen to be a four-dimensional sphere or torus. This 
state of affairs is
somewhat ironic, because the possibility of including a sum over
topologies has often been praised as an advantage of the path
integral formulation over canonical quantization methods, which
employ a 3+1 split of spacetime into space and time. 
As we have argued, this is only true at a formal level, that is, as long
as one does not perform concrete computations (and therefore has
to worry about the convergence or otherwise of an expression like
\rf{propreg}). 
At least from a Euclidean point of view, there are now no further
natural restrictions one may impose on the geometries, and it is
from this starting point that 
the original approach of Dynamical Triangulations proceeded
\cite{edt,edt1}, in order to 
study the properties of the theory (hopefully) defined by the continuum 
limit of \rf{propreg}.

This may be a convenient moment to make some 
non-technical remarks
on how (C)DT evaluates the path integral and extracts physical
information from it, such as the expectation values of certain geometric
observables. A direct analytical evaluation of \rf{propreg} -- although 
available in lower-dimensional models -- is formidably difficult. 
However, unlike in a variety of other approaches to quantum gravity,
DT possesses a set of powerful and well-developed numerical tools, 
whose value can hardly be overstated. They have been adapted
from statistical mechanics and the theory of critical phenomena
to the case where the individual configurations are curved 
geometries, rather than spin or field configurations on a fixed
background space or lattice. The ensemble of spacetimes
underlying the path integral is simulated by Monte Carlo methods \cite{barkema},
generating a random walk in the space of all configurations
according to a probability distribution defined by 
\rf{propreg}.\footnote{For the Euclidean path integral, one can
directly use the real weights $\exp {iS^{\rm reg}[T]}$. For the
Lorentzian case of CDT, in order to obtain a probability distribution
from \rf{propreg}, one first has to apply a so-called
``Wick rotation" which converts the complex to real weights \cite{ajl4d} (see
also footnote \ref{boltz}).}
The limitations of the computer imply that this procedure can only be
implemented on a (possibly large but) finite space of geometric
configurations.\footnote{The computer power involved
in obtaining the results reported here amounted to a few PCs and
work stations.} This is usually taken into account by performing
the simulations on the ensemble of triangulations of a fixed 
discrete volume $N_4$. By repeating the numerical measurements
for a variety of different $N_4$'s, one then tries  
to extrapolate the results in a systematic way
to the physically relevant limit $N_4\rightarrow\infty$. This well-known
technique is known as ``finite-size scaling". 

Now, what kind of ``quantum geometry" does one expect to see with the
help of these tools? If all goes well, the quantum superposition
\rf{propreg} of geometries should be able to reproduce a 
classical spacetime at large scales $L$, that is, in the classical limit. 
However, at small
scales $l$, with $a\ll l\ll L$, one expects quantum fluctuations to
dominate, with a resulting highly nonclassical behaviour of the
geometry. 
To cut a long story short, this was unfortunately not what was found for
the Euclidean dynamically triangulated path integral studied in
the 1990s. This was not immediately realized, but emerged gradually
as more numerical simulations were performed \cite{crumple}. It turned out
that the quantum geometry generated by Euclidean DT
could be in either one of two ``phases". In 
the first one the geometry was completely crumpled, and in the other
totally polymerized, that is, degenerated into thin branching 
threads\footnote{See the next section for a geometric characterization
of these phases.}. These structures persist also at large scales, and
as a result the DT path integral appears to have no meaningful
classical limit, and therefore does not satisfy a necessary
criterion for a theory of quantum gravity. (One can only
wonder how long it may have taken to realize this, had one not been
in a state to perform extensive simulations of the model.) 

The starting point of CDT was the hypothesis that this failure may
have to do with the unphysical {\it Euclidean} nature of
the construction, and that one may be able to rectify the situation by
encoding the causal structure of {\it Lorentzian} spacetimes  
explicitly in the choices of building blocks and gluing rules. 
Several years passed since this initial conjecture, in which CDT's
causal quantization program was implemented and its viability
tested in lower dimensions \cite{2dcdt,3dcdt}. Namely,
superpositions like \rf{propreg} can be defined also by considering
spacetimes glued from two- or three-dimensional building blocks.
This gives rise to simplified toy models which share some, but by no 
means all properties of the true CDT path integral. On the plus side,
they can be tackled both analytically and numerically, and
compared with other quantization approaches to Einstein gravity in
two and three spacetime dimensions. These extensive investigations
showed unequivocally that the causal, Lorentzian path integral 
in all cases gave different results from the corresponding Euclidean
path integral. Encouraged by this and a number of further interesting
and new results pertaining to lower-dimensional quantum gravity, the
investigation moved on to the four-dimensional case in 2004. 

\subsection*{Causality implies four-dimensionality!}

We will now describe the first piece of evidence which showed that
CDT can reproduce at least {\it some} aspects of classical geometry
correctly. This concerns a point where previous related
quantization attempts have failed, namely, to generate a geometric
object that can be said to be {\it four-dimensional} on sufficiently large
scales. 

It may come as a surprise that a superposition of locally
four-dimensional geometries can give anything that is {\it not} again
four-dimensional. After all, we have obtained our geometric
building blocks by cutting out small pieces from a four-dimensional flat
space. However, as is illustrated by Euclidean dynamically triangulated 
models, it is perfectly possible that the dimension comes out not as four,
and this is indeed what seems to happen {\it generically}. 
The crumpled and polymeric phases
of the Euclidean model mentioned in the previous section are characterized by
a so-called Hausdorff dimension which takes the values infinity and two 
respectively.\footnote{Interestingly, one observes the same behaviour 
when one starts from three- instead of four-dimensional building blocks.
This result is believed to hold generically in Euclidean DT models,
as long as the dimensionality of the elementary
building blocks is at least three.}

How can one obtain spaces with such strange dimensionalities?
Roughly speaking, the Hausdorff dimension is obtained by comparing
the typical linear size $r$ of a convex subspace of a given space
(e.g. its diameter) with its volume
$V(r)$. If the leading behaviour is $V(r)\sim r^{d_H}$, the space is
said to have the Hausdorff dimension $d_H$. To obtain an effectively
infinite-dimensional space from gluing $N_4$ four-dimensional simplices
with edge length $a$, one may consider a sequence of triangulations 
whose volume goes to infinity, 
$N_4\rightarrow\infty$, where the gluing for each fixed $N_4$ 
is chosen such that every
single building block shares a given vertex. That is, no matter how
large $N_4$ gets, all building blocks of the triangulated space
crowd around a single point.
This is a procedure which is compatible with the gluing rules, but 
gives rise to a space whose dimensionality diverges,
simply because its linear size always stays at the cutoff length $a$.
Conversely, one can get an effectively one-dimensional space by
gluing the four-dimensional building blocks into a long and thin tube.
That is, as $N_4\rightarrow\infty$ and $a\rightarrow 0$, 
one keeps three out of the four
directions at a size of the order of the cutoff $a$, and only grows the geometry
along the fourth direction.

This argument shows that there are spaces with ``exotic" dimensionality
which can be obtained as limiting cases of regular simplicial
manifolds. Of course, the relevant question for the gravitational path
integral is whether geometries of this nature indeed {\it dominate
the path integral in the continuum limit}. This is a genuinely dynamical
question which cannot be decided a priori. It depends on the relative
weight of ``energy" and ``entropy", that is to say, the Boltzmann weight
of a given geometry (which in turn is a function of the values of the bare
coupling constants) {\it and} the number of geometries with a given,
fixed Boltzmann weight. Thus it may happen that an exotic geometry
(for example, one of the highly crumpled objects above)
has a very large Boltzmann weight and is therefore ``energetically
favoured", but that there are relatively speaking far fewer of such
objects in the ensemble than there are of the more ``normal" geometries,
such that the contribution of the former will in the end play no role in the
path integral in the continuum limit. As we have seen, this is not what
happens in Euclidean dynamically triangulated models for quantum
gravity whose state sums, depending on the values of the coupling constants,
are dominated by exotic geometries which are either maximally
crumpled ($d_H =\infty$) or of the form of so-called branched polymers
(with $d_H =2$). 

The finding that ``dimensionality" is turned into a dynamical 
quantity is
a consequence of the fact that the nonperturbative gravitational path
integral contains highly nonclassical geometries which are curved
(and even highly curved) at the cutoff scale $a$. It can and indeed does
happen that geometries with such an unruly short-scale behaviour
dominate the path integral as $a\rightarrow 0$. As already remarked
earlier, this is exactly what one would expect in analogy with
the path integral for the particle, which in the continuum limit
is dominated by totally nonclassical paths with ``infinitely many corners".
It is important to emphasize that the short-scale picture of geometry
that arises in CDT is completely different from that of the classical theory.
If one looks at a piece of classical spacetime -- no matter how curved --
with an ever finer resolution, it will {\it always} eventually start looking 
like a piece of flat spacetime, namely, when the observed scale
becomes much smaller than the characteristic scale at which 
the space is curved. By contrast, a typical ``quantum spacetime" 
generated by our nonperturbative path integral construction will
{\it never} resemble a flat space, no matter how fine we choose the
resolution of our virtual magnifying glass.

Having understood that quantum geometry will
necessarily 
look very nonclassical at short scales, we presumably are still
left with many
possibilities for the precise microstructure that is generated by various
prescriptions for setting up the gravitational path integral. Can we
formulate criteria for recognizing when a particular 
prescription stands a chance of leading to the correct theory of
quantum gravity? Fortunately, the answer is yes, and the criteria in question
have to do with reproducing features of classical geometry at sufficiently
large scales. As alluded to above, the simplest such test is whether the
quantum geometry has the correct dimension four at large
distances. A path integral which does not pass this test simply does
not qualify as a candidate for a theory of quantum gravity.

The art is then to come up with a path integral which 
allows for large short-scale fluctuations in curvature, but 
in such a way that the
resulting large-scale geometry nevertheless does not degenerate 
completely, so that a sensible classical limit may exist. 
The method of causal dynamical triangulations has for the first
time in the history of the nonperturbative gravitational path integral
given us an explicit example of such a geometry. 
What has been found to be crucial in its derivation are
certain causal rules one imposes on the triangular building blocks,
which make explicit reference to the Lorentzian structure of the
individual geometries contributing to the path integral. 
The new and
intriguing physical insight that can therefore be deduced from this result
is that causality at sub-Planckian scales may be responsible for the 
fact that our universe is four-dimensional \cite{ajl-prl,universe}. A related lesson
that has been made explicit by the dynamical triangulations 
approach in general is the fact that once geometric excitations
are ``let loose" in a nonperturbative formulation of quantum gravity,
just about anything can happen. Not even the dimensionality of
(what we thought of as) the spacetime emerging from  
the quantum superposition has to come out right. At the same time one
could therefore also worry that other nonperturbative quantum gravity
approaches may suffer related pathologies, 
which have only gone undetected because one has not been able
to determine expectation values like that of the Hausdorff dimension
$\langle d_H\rangle$ explicitly.

\begin{figure}[h]
\psfrag{t}{{\bf{\LARGE $\tau$}}}
\centerline{\scalebox{0.4}{\rotatebox{0}
{\includegraphics{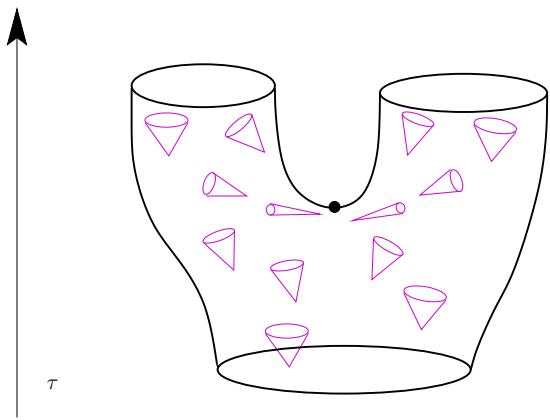}}}}
\caption[trousers]{\small A ``trouser" spacetime as example of a 
spacetime with topology change. Two-dimensional 
rendering of a space (here depicted as
a one-dimensional circle) splitting into
two components as time $\tau$ progresses. The smooth assignment of
light cones must break down at the marked branching point,
because the light cone ``does not know where to turn".}
\label{trousers}
\end{figure}

The reader may by now be curious about the precise nature of the 
causality conditions present in the CDT approach. They are simply
that each spacetime appearing in the sum over geometries
should have a specific form. 
Namely, it should be a geometric object
which can be obtained by evolving a purely spatial geometry in time, in
such a way that its spatial topology (the way in which space hangs together)
is unchanged as a function of time. An example of a forbidden spacetime
is one where an initially connected space splits into two or several
components, or the converse process where several components of
a space reunite into a single one \cite{dowker}. 
Spacetimes with so-called wormholes
also fall into this category and will therefore not be included in the sum 
over geometries.  So, what is wrong with these geometries? Why do
they violate causality? Let us start by explaining why these geometries
are pathological {\it from a classical point of view}. Imagine a 
three-dimensional space that undergoes a branching process as time
progresses (Fig.\ \ref{trousers}). 
Initially the space consists of a single piece (or component),
which simply means that any point in the space can be reached from
any other point along a continuous path. 
At some moment in time, the space splits into two components
which then remain cut off from each other. Classically this represents a
highly singular process, with nothing to suggest such processes
actually occur in nature.\footnote{If this were the case, we would see whole chunks
of space (together with their contents) suddenly disappear.} From a
spacetime point of view, something in these processes goes wrong with the light cone
structure. The assignment of light cones to spacetime
points cannot be smooth, since there must be at least one point in spacetime
(precisely the branching point) where it cannot be decided whether
a light ray arriving from the past should be continued to the future in one or the
other of the two spatial components. Since the light cones define the
causal structure of spacetime, this is an example of a geometry 
where causality is violated. 
The classical Einstein equations simply cannot describe such
topology-changing spacetimes. Two more things should be noted:
firstly, the absence of branching points (and their time inverses, the 
joining points) from a Lorentzian geometry
is invariant under diffeomorphisms because
different notions of time always share the same overall direction of
their ``time flow".\footnote{We are not considering the possibility of a
complete reversion of the time flow, which exchanges past and
future globally.} In order to introduce branching points and their associated
``baby universes" (those parts of the universe that branch out from
the ``mother universe", never to return), one would need to reverse the time flow in entire
open regions of spacetime, which cannot be done by an allowed coordinate
transformation. Secondly, in the Euclidean theory, which has
no distinguished (class of) time direction(s), one simply cannot talk about the
absence or presence of analogous branching points in a meaningful (i.e.
coordinate-invariant) manner.   

Returning to our discussion of the {\it quantum} theory, the premise of
CDT is to use the Lorentzian structure of its contributing geometries
explicitly and exclude all spacetimes with topology changes and
therefore acausal behaviour.\footnote{It is rather straightforward to
implement the causality conditions on the triangulated geometries
of CDT. Each spacetime is built from layers of fixed duration 
(one ``length step" in proper time), and one implements gluing
rules for the simplices which ensure that no change of spatial topology 
can occur during the step \cite{ajl4d}.} Although classical considerations of
causality have motivated a similar implementation of causality in CDT,
it should be emphasized that such constraints on the path integral histories 
can never be derived conclusively from the classical theory. After all,
the individual path integral geometries are never going to be 
smooth classical
objects (let alone solutions to the equations of motion), nor even
close to classical geometries. There is hence no obvious reason to forbid
any particular quantum fluctuations of the geometry, including those
that include topology changes. In principle, a quantum superposition
of acausal spacetimes could lead to a quantum spacetime where causality
by some mechanism is restored dynamically, at least macroscopically. 
However, although this is a theoretical possibility, it is not what one has
observed in the Euclidean version of DT which does not have such causality 
restrictions, and which goes wrong already in trying to
reconstruct a {\it four}-dimensional space. 
By the same token, the fact that individual path integral geometries in CDT
are causal is also not by itself sufficient to guarantee that the quantum geometry
it generates has again the same property. Whether this is indeed the
case is not yet known, and requires a more detailed knowledge
about the local geometric structure than is currently available. 
For example, one would like to ascertain that at a
sufficiently coarse-grained level the quantum geometry possesses a
well-defined light cone structure by defining and measuring suitable quantum 
observables. -- However, there are already a number of important facts 
known about the large- and small-scale structure of the quantum spacetime 
emerging in CDT from first principles, which form
the subject of the following section.

\subsection*{What {\em is} the quantum spacetime generated by CDT?}

We have emphasized in the last section the importance of the emergence
of classical geometry as a test for potential quantum gravity theories.
The dimensionality of spacetime is only one of many quantum
observables one may try to evaluate in order to determine the 
properties of the ground state geometry generated by CDT at various
length scales. It is the coarsest such variable, because the dimensionality
of a spacetime -- at least in classical differential geometry -- precedes
that of specifying a metric structure. 

Talking about observables, one must keep in mind
that an innocent-looking question like ``what is the value of the metric 
tensor $g_{\mu\nu}$ at point $x$?" is among the most difficult to answer
in a nonperturbative approach like ours. Firstly, 
although CDT histories come with a notion of proper time, they do not
otherwise carry any natural coordinate system. Even if we introduced coordinate
systems on the individual triangulated spacetimes, there is no way to
mark ``the same point" simultaneously in all of them. This is a consequence
of the fact that
individual points do not have any physical significance in empty space;
in the absence of matter there is simply nothing we could ``mark" the
point $x$ with. 
We are thus forced to phrase any question about local curvature
properties, say, in terms of quantities that {\it are} meaningful in the context of
a diffeomorphism-invariant theory, for example, $n$-point correlation 
functions where the
location of each of the $n$ points has been averaged over 
spacetime.\footnote{Two-point functions of this type have been measured 
previously in Euclidean DT \cite{correl}.} 

The correlation function that has
been studied up to now in CDT measures the correlation between the
volumes $V_{\rm space}(\tau )$ 
of spatial slices (slices of constant time $\tau$)
some fixed proper-time distance $\Delta \tau$ apart, that is,
a suitably normalized version of the expectation value
\begin{equation}
\langle  V_{\rm space}(0) V_{\rm space}(\Delta \tau)\rangle =\sum_{\tau=0}^{t}
\langle V_{\rm space}(\tau ) V_{\rm space}(\tau+\Delta \tau)\rangle ,
\label{volcorr}
\end{equation}
where the ensemble average is taken over simplicial spacetimes 
with time extension $t$ and with fixed four-volume \cite{ajl-prl,semi,universe}.
One piece of evidence for the four-dimensionality of spacetime
at large distances is the fact that in order to map the functions 
$\langle  V_{\rm space}(0) V_{\rm space}(\Delta \tau)\rangle$ on top of 
each other for different values of the spacetime volume $N_4$, 
the time distance $\Delta \tau$ has to be rescaled by the power $N_4^{1/D_H}$,
where the ``cosmological Hausdorff dimension" is 
$D_H\equ 4$ within measuring accuracy \cite{ajl-prl,universe}. This means that
what we would like to call a continuum 
``time" really scales with the correct fraction of the total spacetime volume. 
Such a
``canonical scaling" is what one would have expected na\"\i vely, but 
is absolutely not ensured a priori in the presence of large geometric
quantum fluctuations, even though the individual building blocks
at the cutoff scale are four-dimensional.\footnote{Further, 
independent evidence that the volumes
$V_{\rm space}(\tau)$ of the spatial slices also scale canonically as $N_4$ is
increased, $V_{\rm space}\sim N_4^{3/D_H}$, with $D_H\equ 4$,
can be found in \cite{universe}.}

Before looking at another striking result on dimensionality to have come
out of CDT, let us review what else we know about the large-scale geometry
of the quantum spacetime dynamically generated by CDT. This concerns a result
which enables us to make contact with (quantum) cosmology. Recall the
remarkable fact that almost every aspect of today's standard model of cosmology,
describing the large-scale structure of our universe, is
based on a radical truncation of (the geometric sector of) 
Einstein's theory to a {\it single} global degree of freedom, the so-called {\it scale
factor} $a(\tau)$. It describes the linear size (or ``scale") of the universe as
a function of time $\tau$.\footnote{As far as we can tell, our present universe not only
expands, but expands at an ever increasing rate, that is, both $\dot a(\tau)> 0$
and  {\it \"a}$(\tau)> 0$. 
A ``big bang" or ``big crunch" corresponds classically to a singular point
where $a\equ 0$.} This truncation is justified if the universe is homogeneous
and isotropic at the largest scales, which means that it looks the same
everywhere and in all spatial directions, something that is usually assumed
to be true. An entirely different question is whether one can extract 
information about the {\it quantum} behaviour of the universe (for example,
very close to the big bang where quantum effects should come into play)
by quantizing the classically truncated system of just a single geometric
variable $a(\tau )$. One may wonder whether in this way one is not missing 
important physics contained in the quantum fluctuations
of all the local gravitational degrees of freedom which the cosmological
description ignores. 

\begin{figure}[ht]
\centerline{\scalebox{0.4}{\rotatebox{0}{\includegraphics{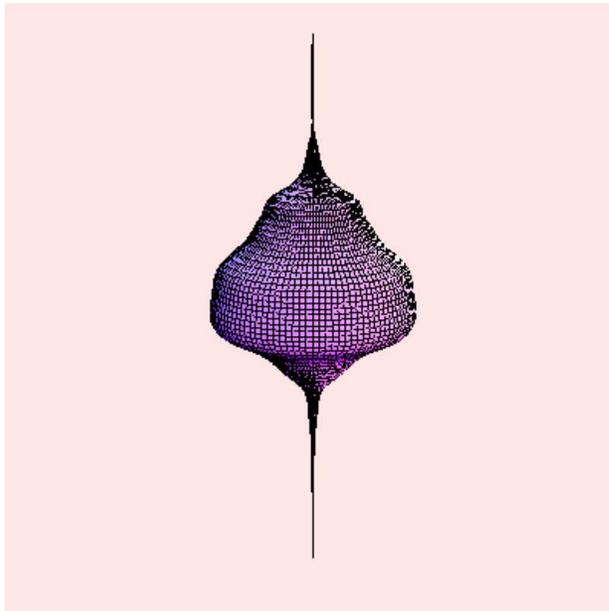}}}}
\caption[phased]{{\small
The overall shape of the extended quantum universe generated by CDT is 
determined by a bounce solution of the Euclidean effective action for the
scale factor $a(\tau)$. The figure is a Monte Carlo snapshot of a typical
universe of $N_4\equ$ 91.100 four-simplices and illustrates the behaviour 
of $a(\tau)$ (the circumference of the rotational body)
as a function of the proper time $\tau$ (vertical axis).}}
\label{unipink}
\end{figure}

Having in hand an explicit construction of quantum geometry \`a la CDT 
where no such truncation is present, one can ask what predictions it
makes for the dynamics of the scale factor, and compare those to 
standard quantum cosmology. The answer obtained is intriguing: it is
indeed possible to extract an effective action for the scale factor from CDT by
integrating out all other degrees of freedom in the full quantum theory.
The resulting action takes the {\it same} functional form as the standard action of a
``minisuperspace" cosmology for a closed universe, 
{\it up to an overall sign} \cite{semi}. The collective effect of the local gravitational 
excitations seems to result in the same kind of contribution as that 
coming from the scale factor itself,
but with the opposite sign. One way to understand this from an
analogous continuum point of view may be in terms of
so-called Faddeev-Popov determinants, which contribute to the
effective action as a result of gauge-fixing \cite{loll}. The potentially
far-reaching consequences of this result for quantum cosmology are
currently being explored. What has already been established is
that the computer-generated quantum geometry can in the
semiclassical approximation be understood as a so-called 
``bounce", a particular type of solution to the Euclidean equation of 
motion (see Fig.\ \ref{unipink}).
On the basis of this, the infamous ``wave function of the universe" 
$\Psi_0(a)$ \cite{hh,vilenkin} has been computed in CDT as a function of the
scale factor $a$ \cite{semi}. 

However, what is also clear from the computer simulations is that
the semiclassical approximation is no longer an adequate
description of the observed behaviour of the scale factor when
the latter becomes small. This is of course to be expected and is an
indicator for new quantum-gravitational effects appearing at short 
distances. Having gathered some nontrivial evidence that CDT's
quantum geometry reproduces well-known aspects of classical
general relativity on sufficiently large scales, the main focus of
research has to be on what the actual {\it quantum modifications} 
of this structure are. This is the place where new
quantum physics will appear, and our effort will go into describing
it in both a qualitative and quantitative manner.

\begin{figure}[t]
\psfrag{X}{{$\sg$}}
\psfrag{Y}{{ $D_S$}}
\centerline{\scalebox{1.1}{\rotatebox{0}{\includegraphics{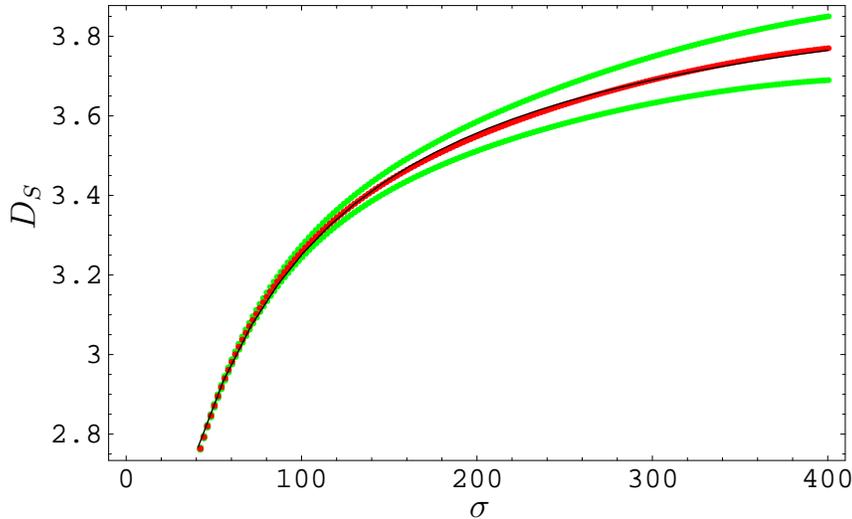}}}~~~~
~~~}
\caption[phased]{{\small The spectral dimension $D_S$ of the universe as
function of the diffusion time $\sg$, as measured in CDT for a quantum universe
of $N_4\equ 181.000$ building blocks. Extrapolating from this data, one obtains
the estimates $D_S(\sg\equ \infty) = 4.02 \pm 0.1$ in the limit of large distances
and $D_S(\sg\equ 0)= 1.80 \pm 0.25$ for the short-distance spectral dimension
\cite{spectral,universe}. The two outer curves represent error bars.}}
\label{spectralfinal}
\end{figure}

CDT has already given us first insights into what the microstructure of 
quantum spacetime may be. The evidence comes from yet another
way of probing the effective dimensionality of spacetime. The idea
is to define a diffusion process (equivalently, a random walk) on the
triangulated geometries in the path integral over spacetimes, and
to deduce geometric information of the underlying quantum
spacetime from the behaviour of the diffusion as a function
of the diffusion time $\sigma$ inherent to the process. 
The beauty of this procedure is its
wide applicability, since diffusion processes cannot just be defined
on smooth manifolds, but on much more general spaces, such as
our triangulations and even on fractal structures \cite{bah}. The quantity we are
interested in is the so-called ``spectral dimension", which is really
the effective dimension of the carrier space ``seen" by the diffusion
process. It can be extracted from the
return probability $P(\sigma)$ which measures the probability of
a random walk to have returned to its origin after diffusion time 
$\sigma$ (or $\sigma$ evolution steps if the diffusion is implemented
discretely).
For diffusion on a flat $d$-dimensional manifold, we have the
exact relation $P(\sigma)=1/(4\pi\sigma)^{d/2}$. For general
spaces we {\it define} the spectral dimension $D_S(\sigma)$ as the
logarithmic derivative\footnote{The complete expression for the return
probability has correction terms because of the finite size
of the computer-generated geometries which we are suppressing
for simplicity. A more detailed discussion can be found in \cite{universe}.} 
\begin{equation}
D_S(\sg) := -2 \;\frac{d\log P(\sg)}{d\log \sg}.
\label{spec}
\end{equation}
Note that in general this dimension will depend on $\sigma$:
small values of $\sigma$ probe the small-distance properties of
the underlying space, and large values its large-distance 
geometry.\footnote{As usual in a random walk, the linear distance
probed will be of the order of $\sqrt{\sigma}$.} The spectral
dimension extracted for the quantum geometry of CDT is a twofold
average over the starting point of the diffusion process (which
is initially peaked at a given four-simplex) and over all geometries
contributing to the path integral \cite{spectral,universe}. The result of the measurement
is quite striking and plotted in Fig.\ \ref{spectralfinal}. 
What one observes is indeed
a scale-dependence of the spacetime dimension! At large distances
it approaches the value four asymptotically, in agreement with the
dimension obtained previously from scaling arguments, and in
agreement with our classical expectation. However, as we probe
the geometry at ever shorter distances (and {\it before} we enter
the region where the simulations become unreliable due to
discretization effects), this dimension decreases continuously
to an extrapolated value of two within measuring accuracy.
Such a scale-dependence has never before been observed in
statistical models of quantum gravity and 
is a clear indication that spacetime behaves highly nonclassically
at short distances close to the Planck length. 
Further investigations of a number of 
critical exponents and dimensions associated with the geometric structure
of spatial slices and ``sandwiches" (of time extension $\Delta \tau=1$)
\cite{universe} suggest the presence of a {\it fractal microstructure} of quantum spacetime,
whose details are the subject of ongoing research. 

In an independent
development, a similar smooth running from four to two of the spectral dimension 
has been derived within a renormalization group approach to quantum
gravity in the continuum \cite{fractal}, which posits (and provides some evidence for) 
the existence of a nontrivial
fixed point in the ultraviolet (i.e. short-distance) regime of quantum gravity
\cite{reuter}. Although this coincidence by no means proves that either 
formulation is correct, it is nevertheless remarkable that the same unexpected
result has been obtained in two very different approaches to quantum gravity. 
If the result can indeed be shown to be part and parcel of a viable quantum
gravity theory, its implications for how we view spacetime and how we
compute quantum processes of the other fundamental interactions on
spacetime may be profound. For example, it could provide a natural 
ultraviolet cutoff for scattering amplitudes in high-energy physics.

\subsection*{Conclusions and outlook}

This article has offered an overview of some of the fundamental
issues addressed by quantum gravity, and has described a particular
attempt to arrive at a consistent quantum theory of gravity, 
through the use of causal dynamical
triangulations (CDT). As we hope to have illustrated, this approach
has yielded a number of concrete results concerning the emergence
of classical geometry from a Feynman-type superposition of spacetimes,
provided appropriate care was taken to eliminate spacetimes with acausal
features from the nonperturbative gravitational path integral. Although the
derivation of the four-dimensionality of spacetime ``from scratch" is an
unprecedented result, more features of the classical theory still need to
be established, for example, the presence of attractive gravitational forces
as expressed by Newton's law. Assuming that this can be accomplished,
the really interesting and new physics lies of course beyond the
classical approximation. Here the challenge will be to extract more
detailed information about the short-scale structure of quantum spacetime
and, if possible, to uncover concrete physical consequences that may
in principle be detectable. As we have seen, CDT offers already some
tantalizing glimpses of what spacetime may look like at or near the
Planck scale. 

It is unlikely that the construction we have presented here will satisfy 
everyone's prejudices of how a quantum theory of gravity should be
constructed, be it through invoking this or that kind of fundamental
discrete structure at the Planck scale or according to this or that 
favourite symmetry principle. This need not necessarily
be a reason for concern: if we can find {\it one} way to Rome, we will
be able to find many others. That is to say, we believe (in the spirit of
``universality") that there is at most one theory which
describes the nontrivial quantum dynamics of intrinsic spacetime 
geometry, and that in order to construct it, we should just get a few
``basic things" right, among them presumably some genuinely nonperturbative
features (like the inclusion of locally highly curved geometries which are
``very far away" from any classical spacetimes), and possibly a principle of
``microcausality", like that implemented in CDT to ensure the emergence
of an extended, four-dimensional spacetime.  

What remains to be shown is that a single such theory with the correct
properties exists. Because of the minimalist input we have used in our
construction (no new symmetries, no new dynamical fields or other
extended objects, no additional spacetime dimensions, and thus no
associated new free parameters) we are unlikely to run into the converse,
``M-theoretic" problem of having vast numbers of possible vacua \cite{vacua}
and therefore possible theories of quantum gravity, with a continuum of
different physical predictions. The paradigm of spacetime beginning
to emerge from CDT is that of a scale-invariant, fractal and effectively
lower-dimensional structure at the Planck scale, which only at a larger scale
acquires well-known features of geometry which accord with our classical
intuition. The deeper reasons for how and why this comes about remain to be 
understood.

\vspace{.7cm}

\noindent {\bf Acknowledgement.} We thank N.\ Mousseau for a critical reading
of the manu\-script. -- This collaboration is part of and supported through
ENRAGE (European Network on 
Random Geometry), a Marie Curie Research Training Network supported by the 
European Community's Sixth Framework Programme, network contract
MRTN-CT-2004-005616. In addition, R.L. acknowledges support by the 
Netherlands Organisation for Scientific Research (NWO) under their VICI program.

\newpage

{\it Jan Ambj\o rn} completed his PhD in
Copenhagen in 1980 on non-perturbative aspects of QCD.
After fellowships at Caltec and Nordita he
moved to the Niels Bohr Institute where he is now a professor.
His research interests include quantum gravity, string theory,
gauge theory and cosmology (in particular questions related
to the baryon asymmetry of the universe), and he has been involved
in numerous analytical and numerical investigations of their properties. 
He is one of the founders of the non-perturbative 
``Dynamical Triangulations" or ``Matrix Model" approach to string theory
and quantum gravity, which provides a diffeomorphism-invariant 
regularization and allows for both analytical solutions and
numerical simulations of these theories. Most recently his research
has been centred around ``Causal Dynamical Triangulations" and the
corresponding theory of quantum gravity. 
\\

{\it Jerzy Jurkiewicz} obtained his PhD in Theoretical Physics at the 
Jagellonian University in Krakow in 1975. His scientific work was initially centred on
lattice gauge theories, with visiting professorships held in Marseille 
(1981), Paris-Sud (1982) and Utrecht (1983/84). His interest in simplicial 
quantum gravity dates from this period, and was followed by extensive 
research work in Krakow and Copenhagen, where he spent several years 
as a visiting professor at the Niels Bohr Institute. His main area of research
is the field-theoretical description of systems with dynamical geometry, 
which besides quantum gravity includes random networks and applications 
in complex financial systems. He is currently head of the Mark Kac Complex 
Systems Research Centre and head of the Department of the Theory of Complex 
Systems of the Institute of Physics, both at the Jagellonian University.
\\

{\it Renate Loll} studied at Freiburg University, the London School 
of Economics, and Imperial College, where she obtained her PhD in Theoretical Physics 
in 1989. She subsequently decided to work on quantum gravity, then as now one of 
the most challenging unsolved problems 
in high-energy physics. After postdoc positions in Bonn, Syracuse, 
Pennsylvania State University and Florence, she became a 
research associate at the Max-Planck Institute for Gravitational Physics in Golm,
where she also held a Heisenberg Fellowship. Since then, she has 
been one of the main proponents of the new ``Causal Dynamical Triangulations"
approach to quantum gravity. In 2001, she joined the Institute for 
Theoretical Physics of Utrecht University, where she
has been Professor for Theoretical Physics since 2005. She is the scientific
coordinator of ENRAGE, the European Network on Random Geometry.

\end{document}